\documentclass[
reprint,
amsmath,amssymb,aps,
]{revtex4-1}

\newtheorem{theorem}{Theorem}
\newtheorem{proposition}{Proposition}
\newtheorem{lemma}{Lemma}
\newtheorem{corollary}{Corollary}
\newtheorem{remark}{Remark}

\def\ba{\begin{array}}
\def\ea{\end{array}}
\def\be{\begin{equation}}
\def\ee{\end{equation}}

\def\ds{\displaystyle}

\def\b{{\bf b}}
\def\c{{\bf c}}

\def\i{{\bf i}}

\def\u{{\bf u}}
\def\v{{\bf v}}

\def\x{{\bf x}}

\def\0{{\bf 0}}
\def\1{{\bf 1}}
\def\2{{\bf 2}}
\def\3{{\bf 3}}
\def\4{{\bf 4}}
\def\5{{\bf 5}}
\def\6{{\bf 6}}
\def\7{{\bf 7}}
\def\8{{\bf 8}}
\def\9{{\bf 9}}

\usepackage{graphicx}\usepackage{dcolumn}\usepackage{bm}\usepackage{amsmath}\usepackage{amscd}\usepackage{amsfonts}\usepackage{extarrows}

\def\bt{\begin{theorem}}
\def\et{\end{theorem}}
\def\bp{\begin{proposition}}
\def\ep{\end{proposition}}
\def\bc{\begin{corollary}}
\def\ec{\end{corollary}}
\def\bo{\begin{proof}}
\def\eo{\end{proof}}
\def\bx{\begin{example}}
\def\ex{\end{example}}
\def\br{\begin{remark}}
\def\er{\end{remark}}
\def\bl{\begin{lemma}}
\def\el{\end{lemma}}

\begin{document}

\preprint{APS/123-QED}
\title{Reconsider HHL algorithm and its related quantum machine learning algorithms}
\author{Changpeng Shao}
\email{cpshao@amss.ac.cn}
\affiliation{KLSC, Academy of Mathematics and Systems Science, Chinese Academy of Sciences\\ Beijing 100190, China}
\date{\today}

\begin{abstract}
HHL quantum algorithm to solve linear systems is one of the most important subroutines in many quantum machine learning algorithms.
In this work, we present and analyze several other caveats in HHL algorithm, which have been ignored in the past.
Their influences on the efficiency, accuracy and practicability of HHL algorithm and
several related quantum machine learning algorithms will be discussed.
We also found that these caveats affect HHL algorithm much deeper than the already noticed caveats.
In order to obtain more practical quantum machine learning algorithms with less assumptions based on HHL algorithm,
we should pay more attention to these caveats.
\end{abstract}

\pacs{Valid PACS appear here}
\maketitle

\section{Introduction}

As the byproducts of Shor's factoring algorithm \cite{shor} and Grover's searching algorithm \cite{grover96},
the quantum phase estimation algorithm \cite{kitaev} and amplitude amplification technique \cite{brassard-Amplitude-Amplification}
play important roles in many quantum algorithms developed in the past.
It is a beautiful application of quantum phase estimation algorithm to Shor's factoring algorithm.
However, other applications like quantum counting \cite{brassard}, eigenvalue estimation of Hermitian matrix \cite{abrams},
HHL algorithm to solve linear system \cite{harrow} appear some ``unclean" parts (here ``unclean" does not means the algorithms are not good,
but means they may contain some restrictions).
The reason to these ``unclean" parts comes from quantum phase estimation algorithm itself. Since quantum phase estimation algorithm is used
to estimate eigenvalue of unitary transformation, which has the expression $e^{\i \theta}$ with $0\leq \theta <2\pi$.
For example, in the case of estimating eigenvalue $\lambda$ of Hermitian matrix $H$. Although $e^{\i Ht}$ is unitary with eigenvalue $e^{\i\lambda t}$,
the argument $\lambda t$ may not lie between 0 and $2\pi$. We should compress $\lambda$ via $t$ into a small number $\lambda t$ that lies in the interval $[0,2\pi)$.
By quantum phase estimation algorithm, we will get a good approximate of $\lambda t$. However, the error will be enlarged by $t$ and so the complexity will also be enlarged by $t$. This means we may cannot get a good approximate of $\lambda$ efficiently.
What worse is that sometimes we even do not know how to choose such an $t$.

Eigenvalue estimation of Hermitian matrix is one central step of HHL algorithm.
The above discussion points out one caveat in HHL algorithm, that is the choice of the compression parameter $t$.
But in HHL algorithm, the authors have assumed the singular values of the coefficient matrix lie between $1/\kappa$ and 1,
where $\kappa$ is the condition number. And so avoided to consider this problem.
However, in any given problem, we should consider this before applying HHL algorithm in order to make less assumptions.

A known fact is that HHL algorithm only returns a good approximate $|\tilde{x}\rangle$ of the quantum state $|x\rangle$ of the exact solution.
However, such a good approximate may not induce a good approximate of the classical solution $x=|x||x\rangle$, where $|x|$ is the 2-norm of $x$.
This is because, if the error between $|\tilde{x}\rangle$ and $|x\rangle$ is $\epsilon$, then the error between $|x||\tilde{x}\rangle$ and $x$ will be $|x|\epsilon$.
This means the error is enlarged by the norm of the classical solution.
Or in other words, the complexity of HHL algorithm is enlarged by the norm of the classical solution. This is another caveat that did not considered in HHL algorithm. The actual effect of above two discovered caveats on the efficiency of HHL algorithm will be discussed in this work.

It is well known that HHL algorithm paves a way to study machine learning by quantum computer
\cite{aaronson}, \cite{biamonte}, \cite{dunjko}, \cite{rebentros14}, \cite{rebentros17}, \cite{schuld}, \cite{wiebe}.
With the rapid developments of applications of HHL algorithm in quantum machine learning.
On one hand, these works provide substantial examples that quantum computer can do better than classical computers.
On the other hand, we should ask what is the practical efficiency of these works due to the restrictions of HHL algorithm?
And under what kind of conditions, can these works achieve high speedup than the classical counterparts?
The second target of this work is to reconsider the influences of the caveats we discovered in HHL algorithm on the efficiency of
several quantum machine learning algorithms:
linear regression \cite{schuld}, \cite{wiebe}, \cite{wang},
supervised classification \cite{lloyd13},
least-square support vector machine \cite{rebentros14}
and Hamiltonian simulation of low rank matrix \cite{rebentros16}.

The structure of this work is as follows:
In section II, we first briefly review the quantum algorithm to estimate eigenvalues of Hermitian matrix, from which we will discuss some other
caveats about HHL algorithm that did not discussed in the past.
%These caveats affect more about the efficiency of HHL algorithm than the original caveats discussed in \cite{harrow}.
Section III is devoted to reconsider the efficiency of several quantum machine learning algorithms that related to HHL algorithm,
which also possess the same problems in HHL algorithm.
%These findings tell us that we should be more careful to apply HHL algorithm to  certain problems in order to obtain more practical quantum algorithms.

\section{Reconsidering HHL algorithm}

When applying HHL algorithm \cite{harrow} to solve linear systems $Ax=b$ (assume $A$ is Hermitian),
people usually talk about the following four caveats \cite{aaronson}, \cite{childs}:

(C1). The condition number $\kappa$ of $A$.

(C2). The Hamiltonian simulation of $e^{\i A t}$.

(C3). The preparation of the quantum state $|b\rangle$.

(C4). The result of HHL algorithm is a quantum state $|x\rangle$ of the solution.

The caveat (C2) can be solved, for example in the case when $A$ is sparse and all its entries are efficiently available \cite{berry}
or when $A$ is low rank \cite{rebentros16}, \cite{lloyd14}.
There are also some methods to resolve the caveat (C3), for example in the relatively uniform case \cite{lloyd13},
or when all the entries and the norm of $b$ are efficiently computable \cite{grover}.
As for (C4), obtaining quantum state $|x\rangle$ is enough for many quantum machine learning problems
\cite{rebentros14}, \cite{rebentros17}, \cite{schuld}, \cite{wiebe}.
The influence of consider number is unavoidable in the algorithm. From these points, these four caveats seem acceptable to HHL algorithm.
However, in the original paper of HHL algorithm, it actually contains another caveat,
which is related to the first caveat and is easy to be ignored when using HHL algorithm:

(C5). The singular values of $A$ lie between $1/\kappa$ and 1.

The first four caveats may be solvable in some sense as discussed above, however, the fifth caveat is a little difficult to overcome.
As discussed in the original paper \cite{harrow}, it can be solved by a scaling.
But usually, it is hard to achieve such a scaling, since we do not know the condition number $\kappa$ in advance.
The caveat (C5) mainly comes from the quantum phase estimation algorithm to estimate eigenvalues of Hermitian matrix.
We should compress the eigenvalues of the Hermitian matrix into small numbers that lie between 0 and $2\pi$ before applying quantum phase estimation algorithm.
In the following, we first briefly review this algorithm.

Let $A=(a_{ij})$ be an $M\times M$ Hermitian matrix with an eigenvalue $\lambda$ (unknown) and a corresponding eigenvector $|u\rangle$ (known).
Then $U=e^{\i A t}$ is unitary. Now we suppose $U$ can be efficiently simulated in time $O(t^\gamma \textmd{poly}(\log M)/\epsilon)$
with accuracy $\epsilon$. Consider the following quantum phase estimation algorithm to
estimate the eigenvalue $\lambda$, where $t$ and $N$ appear below are under determination:
\be\ba{lll}\label{phase-estimation-alg0} \vspace{.15cm}
\hspace{-.3cm}\ds\frac{1}{\sqrt{N}}\sum_{x=0}^{N-1}|x\rangle|u\rangle
&\mapsto&\ds\frac{1}{\sqrt{N}}\sum_{x=0}^{N-1}|x\rangle U^x|u\rangle \\ \vspace{.15cm}
&=& \ds\frac{1}{\sqrt{N}}\sum_{x=0}^{N-1}e^{\i\lambda tx}|x\rangle|u\rangle \\
&\mapsto& \ds\frac{1}{N}\sum_{y=0}^{N-1}\Bigg[\sum_{x=0}^{N-1}e^{\i x(\lambda t-\frac{2\pi y}{N})}\Bigg]|y\rangle|u\rangle.
\ea\ee
Then
\be\ba{lll} \vspace{.15cm}
\textmd{Prob}(|y\rangle)
&=& \ds\frac{1}{N^2}\Bigg|\sum_{x=0}^{N-1}e^{\i x(\lambda t-\frac{2\pi y}{N})}\Bigg|^2 \\
&=& \ds\frac{1}{N^2}\Bigg|\frac{\sin \frac{N}{2}(\lambda t-\frac{2\pi y}{N})}{\sin \frac{1}{2}(\lambda t-\frac{2\pi y}{N})}\Bigg|^2.
\ea\ee
If $|\lambda t-\frac{2\pi y}{N}|\leq \frac{\pi}{N}$, then $|\frac{N}{2}(\lambda t-\frac{2\pi y}{N})|\leq \frac{\pi}{2}$, and so
\be
\textmd{Prob}(|y\rangle)
\geq\frac{4}{\pi^2}\frac{1}{N^2}\Bigg|\frac{\frac{N}{2}(\lambda t-\frac{2\pi y}{N})}{\frac{1}{2}(\lambda t-\frac{2\pi y}{N})}\Bigg|^2
=\frac{4}{\pi^2}.
\ee
By choosing suitable $t$ and $N$, we can always find such a $y$ satisfies $|\lambda t-\frac{2\pi y}{N}|\leq \frac{\pi}{N}$.
Therefore $|\lambda-\frac{2\pi y}{tN}|\leq \frac{\pi}{tN}=\delta$ and $N=O(1/t\delta)$. At this time $2\pi y/tN$ will be a good approximate of $\lambda$.
The complexity of this algorithm is
\be\label{complexity0}
O((tN)^\gamma\textmd{poly}(\log M)/\epsilon)=O(\textmd{poly}(\log M)/\epsilon \delta^\gamma),
\ee
here $\delta$ is the accuracy of estimation of eigenvalue $\lambda$.
The above is a brief overview of quantum phase estimation algorithm to estimate eigenvalues of Hermitian matrix.
On the whole, when fixing $t$ and the accuracy $\delta$, we can choose a suitable $N$ such that the algorithm is efficient.

Note that the result $2\pi y/tN$ of algorithm (\ref{phase-estimation-alg0}) is non-negative, so what if $\lambda<0$?
This problem actually does not hard to solve, since we can choose $t$ small enough, such that $|\lambda t|<\pi$.
So if $2\pi y/N>\pi$, then we believe that $\lambda t=2\pi y/N-2\pi=-2\pi (N-y)/N$.
The good point of quantum phase estimation is that we can estimate all eigenvalues of $A$
even without knowing the eigenvectors. The algorithm is almost the same as (\ref{phase-estimation-alg0}) and it forms one central step of HHL algorithm.
However, there exists one problem we should consider beforehand in algorithm (\ref{phase-estimation-alg0}).

\vspace{.3cm}

\textbf{Problem T. How to choose $t$.} It is clear that $t$ should satisfy $|\lambda t|<\pi$ due to $e^{2\pi \i}=1$ and the sign of $\lambda$.
A theoretical choice of $t$ is $t=\pi /|\lambda_{\max}|$, where $|\lambda_{\max}|$ is the maximal singular value of $A$.
There are several different types of upper bounds about $|\lambda_{\max}|$, a few are listed below:
\be\ba{lll}\vspace{.15cm}
& \hspace{-.3cm}\ds\sqrt{\textmd{Tr}(AA^\dagger)},     &~~ \|A\|_1=\max_j \sum_i|a_{ij}|, \\
& \hspace{-.3cm}\ds\|A\|_2=\sqrt{\sum_{i,j}|a_{ij}|^2},&~~ M\|A\|_{\max}=M\max_{i,j}|a_{ij}|.
\ea\ee
The above choices about upper bound will affect the complexity of the total algorithm,
since the above matrix computation are not easy in classical computer or even in quantum computer.
One simple case is when $A$ is $\textmd{poly}(\log M)$ sparse and all the entries of $A$ are bounded by $\textmd{poly}(\log M)$.

\vspace{.3cm}

Now come back to HHL algorithm.
The authors in \cite{harrow} assume that all the singular values of $A$ lies between $1/\kappa$ and $1$.
At this time, $1/\kappa=|\lambda_{\min}|$ equals the smallest singular value of $A$. Under this assumption,
the complexity of solving $Ax=b$ is
\be\label{complexity-hhl-0}
O((\log M)s^2\kappa^2/\epsilon)=O((\log M)s^2/\lambda_{\min}^2\epsilon),
\ee
where $s$ is the sparseness of matrix $A$.
Moreover, as discussed in the original paper, the assumption (C5) can be resolved by scaling the the linear system in the first step.
However, scaling does not affect the whole procedure of HHL algorithm, since the initial state of HHL algorithm is $|0\rangle|b\rangle$,
and scaling does not affect this initial state. Actually, scaling only works when estimating the eigenvalues of $A$.
It compresses the singular values of $A$ into the interval $[1/\kappa,1]$.
More specifically, let $\widetilde{A}$ be a new matrix which does not satisfy the condition (C5). Denote $A=t\widetilde{A}$ such that
$A$ satisfies the condition described in (C5) (here we suppose $t$ can be obtained by some methods). Then from (\ref{complexity-hhl-0}),
if $|\tilde{\lambda}_{\min}|$ is the smallest singular value of $\widetilde{A}$, the complexity of HHL algorithm will be
\be\ba{lcl}\label{complexity-hhl-1} \vspace{.2cm}
\ds O\Big(\frac{(\log M)s^2}{t^2\tilde{\lambda}_{\min}^2\epsilon}\Big)
&\xlongequal[]{\textmd{theoretically}}& \ds O\Big(\frac{(\log M)s^2\tilde{\lambda}_{\max}^2}{\tilde{\lambda}_{\min}^2\epsilon}\Big) \\
&=& \ds O\Big(\frac{(\log M)s^2\kappa^2}{\epsilon}\Big).
\ea\ee
Here ``theoretically" means the best choice about $t$ is $O(1/|\tilde{\lambda}_{\max}|)$, where $|\tilde{\lambda}_{\max}|$ is the largest singular value of $\widetilde{A}$.
It also means we may cannot obtain the best choices of $t$. Formula (\ref{complexity-hhl-1}) implies that, if we do not know $\kappa$ in advance, and just choose a reasonable $t$ based on some methods, then the smaller $t$ is, the higher complexity of HHL algorithm will be.

Another point in (\ref{complexity-hhl-1}) we should pay attention to is that HHL algorithm only preserves eigenvalues larger than $1/\kappa=1/|\lambda_{\min}|$.
However, we do not know $|\lambda_{\min}|$ beforehand.
There are lots of upper bounds about the maximal eigenvalue, but few about the smallest eigenvalue.
A reasonable idea is choosing a small number $\mu$ instead of $|\tilde{\lambda}_{\min}|$ (for example, see applications in \cite{rebentros14}, \cite{rebentros17}).
Then in the procedure of HHL algorithm,
we only keep the eigenvalues larger than $\mu$ and ignore all the eigenvalues smaller than $\mu$.
At this time, the complexity will be
\be\label{complexity-hhl-2}
O((\log M)s^2/t^2\mu^2\epsilon).
\ee
However, it will make the solution less accuracy if $\mu$ is not close to $|\tilde{\lambda}_{\min}|$.
In other words, if we only consider the solution of the linear system lie in these components, then the error is $\epsilon$.
While the error to the exact solution may be larger than $\epsilon$. Since the solution not only depends on $A$, but also on $b$.
And it seems hard to estimate the error of HHL algorithm at this time.
This can be reflected more clearly in the simple case when $A=\textmd{diag}\{a_0,\ldots,a_{r-1},a_r,\cdots,a_{M-1}\}$ is a diagonal matrix
with
\[1\geq|a_0|\geq \cdots\geq|a_{r-1}|\geq \mu>|a_r|\geq \cdots\geq|a_{M-1}|>0.\]
Set $b=(b_0,\ldots,b_{r-1},b_r,\ldots,b_{M-1})^T$.
Then in HHL algorithm, we will obtain a solution in the form
\[(b_0/a_0,\ldots,b_{r-1}/a_{r-1},0,\ldots,0)^T.\]
However, the exact solution is
\[(b_0/a_0,\ldots,b_{r-1}/a_{r-1},b_r/a_r,\ldots,b_{M-1}/a_{M-1})^T.\]
The error will be large if $b$ does not lie in the well-conditioned parts of $A$.

The final point we should pay attention to HHL algorithm is that: HHL algorithm returns an approximate state $|\tilde{x}\rangle$
of the state $|x\rangle$ of the exact solution. This means $||x\rangle-|\tilde{x}\rangle|\leq \epsilon$. But the exact solution is $x=|x||x\rangle$. So
$|x-|x||\tilde{x}\rangle|\leq |x|\epsilon$. The error may be enlarged by the norm $|x|$. If we want this error small, then the complexity of HHL algorithm becomes
$O((\log M)s^2\kappa^2|x|/\epsilon)$. Actually, this phenomenon has already appeared in the quantum counting algorithm \cite{brassard}: Suppose there are $K$ marked items in $\{1,2,\ldots,N\}$. Then the result of quantum counting algorithm is that we can approximate $K$ with relative error $\epsilon$ in time $O(\sqrt{N/K\epsilon^2})$.
So we can find an $\widetilde{K}$ such that $|K-\widetilde{K}|\leq K\epsilon$. Similarly, if we want $K\epsilon=1$ for example, then the final complexity
will be enlarged into $O(\sqrt{NK})$.

The above phenomenon in HHL algorithm can be stated more clearly in the following way:
Assume $A$ is invertible, the eigenvalues of $A$ are $\lambda_1,\ldots,\lambda_M$ and the corresponding eigenvectors are $|u_1\rangle,\ldots,|u_M\rangle$.
Assume $b=\sum_{j=1}^M\beta_j|u_j\rangle$, then the exact solution of $Ax=b$ is $x=\sum_{j=1}^M \beta_j\lambda_j^{-1}|u_j\rangle$. And so its quantum state is
$|x\rangle=\frac{1}{\sqrt{Z}}\sum_{j=1}^M \beta_j\lambda_j^{-1}|u_j\rangle$, where $Z=|x|^2=\sum_{j=1}^M |\beta_j\lambda_j^{-1}|^2$. Assume the solution obtained
by HHL algorithm is $|\tilde{x}\rangle=\frac{1}{\sqrt{\widetilde{Z}}}\sum_{j=1}^M \beta_j\tilde{\lambda}_j^{-1}|u_j\rangle$
where $\widetilde{Z}=\sum_{j=1}^M |\beta_j\tilde{\lambda}_j^{-1}|^2$ and $|\lambda_j^{-1}-\tilde{\lambda}_j^{-1}|\leq\epsilon$.
Then the classical solution obtained by HHL algorithm has the form $\tilde{x}=\sum_j \beta_j\tilde{\lambda}_j^{-1}|u_j\rangle$.
Note that $\widetilde{Z}=|\tilde{x}|^2$.
By the error analysis in HHL algorithm \cite{harrow}, we have
\[O(\epsilon^2) \geq ||x\rangle-|\tilde{x}\rangle|^2=2-2\langle x|\tilde{x}\rangle.\]
This means $1\geq \langle x|\tilde{x}\rangle\geq 1-O(\epsilon^2)$.
On one hand,
\[|Z-\widetilde{Z}|=\sum_{j=1}^M |\beta_j|^2\Big||\lambda_j^{-1}|^2-|\tilde{\lambda}_j^{-1}|^2\Big|\leq O(\epsilon\kappa|b|^2).\]
We wish $\epsilon\kappa|b|^2\leq\epsilon_1$ is small. So $\epsilon\leq\epsilon_1/\kappa|b|^2$.
On the other hand, if we set $\widetilde{Z}=Z+\delta$ with $0\leq\delta\leq \epsilon_1$, then
\[\ba{lll} \vspace{.2cm}
|x-\tilde{x}|^2
&=& \widetilde{Z}+Z-2\sqrt{Z\widetilde{Z}}\langle x|\tilde{x}\rangle \\ \vspace{.2cm}
&\leq& 2Z+\delta-2\sqrt{Z(Z+\delta)}(1-O(\epsilon^2)) \\ \vspace{.2cm}
&\leq& \delta+(Z+\delta)O(\epsilon^2) \\
&\approx& O(Z\epsilon^2).
\ea\]
Similarly, we also wish $Z\epsilon^2\leq \epsilon_2^2$ is small, which implies $\epsilon\leq\epsilon_2/\sqrt{Z}=\epsilon_2/|x|$. Combing the above analysis,
we can choose $\epsilon=\min\{\epsilon_1/\kappa|b|^2,\epsilon_2/|x|\}$. Finally, we assume $\epsilon_1=\epsilon_2=:\tilde{\epsilon}$ when they are small. Then
\be
\epsilon=\min\{\tilde{\epsilon}/\kappa|b|^2,\tilde{\epsilon}/|x|\}
=\frac{\tilde{\epsilon}}{\max\{\kappa|b|^2,|x|\}}.
\ee
The complexity of HHL algorithm will be
\be
O\Big(\frac{(\log M)s^2\kappa^2\max\{\kappa|b|^2,|x|\}}{\tilde{\epsilon}}\Big).
\ee
So the complexity of HHL algorithm is influenced by the norm of the solution and also the norm of $b$.
It is not hard to check that the error between $A\tilde{x}$ and $b$ is
$|A\tilde{x}-b|^2=\sum_{j=1}^M |\beta_j(\lambda_j\tilde{\lambda}_j^{-1}-1)|^2=O(\widetilde{Z}\epsilon^2)$,
which also implies the influence of the norm of the solution on the efficiency of HHL algorithm.

\section{Reconsidering some quantum machine learning algorithms}

The problems discussed in HHL algorithm actually appears in some other related quantum machine learning algorithms. This section will be denoted to
review several of them.

\emph{Linear regression.}
Linear regression is a basic problem in machine learning.
The quantum algorithm to linear regression problem has been considered in \cite{schuld}, \cite{wiebe}, \cite{wang}.
It is well known that the linear regression problem is equivalent to solve a linear system $F^\dagger F \x=F^\dagger \b$, where $F$ is the data matrix and $\b$ is a given vector.
The prediction on the new data $\c$ is equivalent to evaluate the inner product $\c\cdot\x$. By HHL algorithm, we can find the state $|x\rangle$ of the solution efficiently.
And suppose we can prepare the quantum state $|c\rangle$ of $\c$ efficiently. Then by swap test, we can estimate $\langle c|x\rangle$ efficiently.
However, there appears at least two problems. First, as discussed above, the accuracy of HHL algorithm is related to $|\x|$, which may kill the exponential speedup to this problem.
Second, note that $\c\cdot\x=|\c||\x|\langle c|x\rangle$, so a good approximate of $\langle c|x\rangle$ does not imply a good approximate of $\c\cdot\x$ especially when
$|\c|,|\x|$ are large. Therefore, generally swap test is not good to estimate inner product of classical vectors even though their quantum states can be prepared efficiently.

\emph{Supervised classification.}
In paper \cite{lloyd13}, Lloyd et al. provided an efficient quantum algorithm to one type of supervised classification problem.
Such a classification is based on the comparison of distances between the given vector $\u$ to the means of two clusters $V$ and $W$. The main techniques used in this paper are swap test and
quantum state preparation. This paper introduced a great technique to prepare the desired quantum state. More specifically, assume $V=\{\v_1,\ldots,\v_M\}$, then
based on Hamiltonian simulation, they get the following state efficiently
\[\ba{lll}
&& \ds \frac{1}{\sqrt{2}}|0\rangle\Big[\cos(|\u|t)|0\rangle-\frac{1}{\sqrt{M}}\sum_{j=1}^M\cos(|\v_j|t)|j\rangle\Big] \\
&& \hspace{.5cm} \ds -\frac{\i}{\sqrt{2}}|1\rangle\Big[\sin(|\u|t)|0\rangle-\frac{1}{\sqrt{M}}\sum_{j=1}^M\sin(|\v_j|t)|j\rangle\Big].
\ea\]
Choose $t$ so that $|\u|t,|\v_j|t\ll 1$, then the state along with $|1\rangle$ is an approximate of
\be \label{SC-eq1}
\frac{1}{\sqrt{Z}}\Big[|\u||0\rangle-\frac{1}{\sqrt{M}}\sum_{j=1}^M|\v_j||j\rangle\Big],
\ee
where $Z=|\u|^2+\frac{1}{M}\sum_{j=1}^M|\v_j|^2$.
Based on swap test, we can get a good approximate of the probability, which is about $Zt^2$, of getting $|1\rangle$.
It is not hard to see that this technique works well when the data $\{|\u|,|\v_1|,\ldots,|\v_M|\}$ are relatively uniformly distributed due to the choice of $t$.
And the complexity is affected by $\max\{|\u|,|\v_1|,\ldots,|\v_M|\}$. So it requires the norms of given vectors are relatively small.

When obtaining the state (\ref{SC-eq1}), we can measure the first register of the following state
\[\frac{1}{\sqrt{2}}\Big(|0\rangle|u\rangle+\frac{1}{\sqrt{M}}\sum_{j=1}^M|j\rangle|v_j\rangle\Big)\]
in the basis obtained by extending the state (\ref{SC-eq1}). The probability of getting (\ref{SC-eq1}) is
\be
P=\frac{1}{2Z^2}\Big|\u-\frac{1}{M}\sum_{j=1}^M\v_j\Big|^2,
\ee
which can also estimated efficiently by swap test.
Finally, we will obtain a good approximate of the distance, which equals $2PZ^2$, from $\u$ to the mean of $V$.
We should note that although the error of estimating $P$ is small, after multiplying $Z^2$, the error of estimating $2PZ^2$ will be large.
So to make the final error small, the complexity of estimating this distance will be enlarged by $Z$ too.
The complexity to this classification problem analyzed in \cite{lloyd13} is $O(\epsilon^{-1}\log(MN))$, here $N$ is the dimension of the vectors. But based on the above analysis, the complexity should be $O(Z^2\epsilon^{-1}\log(MN))$.

\emph{Least-square support vector machine.}
As discussed in \cite{rebentros14}, the least-square support vector machine problem is equivalent to solve the following linear system
\[F\left(
           \begin{array}{c}
             b \\
             \vec{\alpha} \\
           \end{array}
         \right)=\left(
    \begin{array}{cc}
      0       &~ \vec{1}^T \\
      \vec{1} &~ K+\gamma^{-1}I_M \\
    \end{array}
  \right)\left(
           \begin{array}{c}
             b \\
             \vec{\alpha} \\
           \end{array}
         \right)=\left(
           \begin{array}{c}
             0 \\
             \vec{y} \\
           \end{array}
         \right),\]
where $K$ is the kernel matrix, $\vec{1}^T=(1,\cdots,1)$ and $I_M$ is the identity matrix.
In paper \cite{rebentros14}, Rebentrost et al. provided a quantum algorithm to solve this linear system by HHL algorithm.
A key point of this work is the Hamiltonian simulation of matrix $F$. The main technique comes from the paper \cite{lloyd14}.
They also uses a technique proposed in paper \cite{lloyd13} to estimate the trace of the kernel matrix $K=(\x_i\cdot\x_j)_{M\times M}$,
which is required in the Hamiltonian simulation technique. Based on the method given in \cite{lloyd13}, they got a good approximate of $\textmd{Tr}(K)/M$
in time $\widetilde{O}(1/\epsilon)$. However, a good approximate of $\textmd{Tr}(K)/M$ does not imply a good  approximate of $\textmd{Tr}(K)$.
The error will be enlarged into $M\epsilon$. And in order to make this error small, the complexity becomes $\widetilde{O}(M/\epsilon)$.
So it is a little hard to estimate $\textmd{Tr}(M)$ efficiently. And it seems that this is necessary for them to simulate $e^{\i tF}$.

Actually, the trace of a matrix cannot being efficiently estimated in quantum computer in the general case, otherwise it will contradict the optimality of Grover algorithm.
More specifically, suppose there is a quantum algorithm that can estimate $\textmd{Tr}(B)$ in time $O(\textmd{poly}(\log n)/\epsilon\delta)$
with accuracy $\epsilon$ and probability $1-\delta$ for any $n\times n$ matrix $B$.
In the searching problem, assume $f$ is defined by $f(x)=1$ if $x$ is marked and $f(x)=2$ if $x$ is not marked. Then the trace of the diagonal
matrix with diagonal entries formed by $f(x)$ will be estimated in time $O(\textmd{poly}(\log n)/\epsilon\delta)$ with probability $1-\delta$.
However, the trace of this diagonal matrix corresponds to the summation of all $f(x)$. This means we can decide whether or not there exist marked items in time $O(\textmd{poly}(\log n)/\epsilon\delta)$ with probability $1-\delta$. Then the method of bisection will finally help us find
the marked items in time $O(\textmd{poly}(\log n)/\epsilon\delta)$ with probability $(1-\delta)^{\log n}$. If we choose $\delta=1/\log n$,
then $(1-\delta)^{\log n}\approx1/e\approx 0.368$, which is a constant, here $e\approx2.71828$ is the Euler number. This is impossible.
Even without considering the method of bisection to the search problem,
efficient algorithm to decide whether or not there exists marked items in polynomial time will imply quantum computer can
solve NP-complete problems, which is a highly implausible result \cite{aaronson05}.

\emph{Hamiltonian simulation of low rank matrix.}
In paper \cite{rebentros16}, Rebentrost et al. provided a new method to exponentiate low rank but non-sparse Hermitian matrix.
Instead exponentiating a low rank Hermitian matrix $A$, it exponentiates $A/M$, where $M$ is the size of the matrix.
When considering the above analysis to HHL algorithm (\ref{complexity-hhl-1}),
the choice of $t=1/M$ seems not good to the linear system solving problem,
unless the problem does not related to the eigenvalues but eigenvectors,
just like the Procrustes problem considered in the paper \cite{rebentros16}.
On the other hand, as discussed in \textbf{Problem T}, in the $\textmd{poly}(\log M)$ sparse case and all entries of $A$ are bounded by $\textmd{poly}(\log M)$, then the fifth caveat can be resolved easily. However, in the low rank case, the fifth caveat seems not easy to handle.

As we can seen from the above analysis, the main problem in HHL algorithm and its related quantum algorithms comes from the compression of a large number into a small one.
A good compression brings a lot of helps to the complexity. And this should be done in the first step of the whole algorithm.
But this compression seems not easy to resolve in the general case.
Anyway, HHL algorithm is an important quantum algorithm, and plays an important role in quantum machine learning.
And when applying it to solve other problems, we should be very careful to deal with its restrictions in order to get more practical algorithms.

\textbf{Acknowledgements}. This work was supported by the NSFC Project 11671388, the CAS Frontier Key Project QYZDJ-SSW-SYS022, China Postdoctoral Science Foundation and the Guozhi Xu Posdoctoral Research Foundation.

\nocite{*}
\bibliography{apssamp}

\end{document}